\documentclass[conference]{IEEEtran}

\IEEEoverridecommandlockouts
\usepackage{cite}
\usepackage{amsmath,amssymb,amsfonts}
\usepackage{algorithmic}
\usepackage{graphicx}
\usepackage{textcomp}
\usepackage{xcolor}
\usepackage{comment}
\usepackage{tikz}
\usepackage{orcidlink}
\usepackage{booktabs}
\usepackage{tabularx}
\usepackage{bm}
\usepackage{lipsum} 
\usepackage{caption} 
\usepackage{subcaption}
\usepackage{float}

\ifCLASSINFOpdf
\else
\fi

\def\BibTeX{{\rm B\kern-.05em{\sc i\kern-.025em b}\kern-.08em
    T\kern-.1667em\lower.7ex\hbox{E}\kern-.125emX}}
\begin{document}

\title{
AIaaS for ORAN-based 6G Networks: Multi-time Scale Slice Resource Management with DRL 
}
\author{
    \IEEEauthorblockN{
         Suvidha Mhatre\orcidlink{0000-0003-2601-2774}\textsuperscript{*},
         Ferran Adelantado\orcidlink{0000-0002-9696-1169}\textsuperscript{†},
         Kostas Ramantas\orcidlink{0000-0002-1304-784X}\textsuperscript{‡},
        Christos Verikoukis\orcidlink{0000-0001-8774-1052}\textsuperscript{§}
    }
    \IEEEauthorblockA{
        \textsuperscript{*}\textit{Universitat Politecnica de Catalunya, Barcelona, Spain}
        \textsuperscript{†}\textit{Universitat Oberta de Catalunya, Barcelona, Spain}\\
        \textsuperscript{‡}\textit{Iquadrat Informatica S.L., Barcelona, Spain}
        \textsuperscript{§}\textit{University of Patras and ISI Athena, Greece} \\
    }
    \thanks{This paper was published in ICC 2024, Denver, CO, USA, pp. 5407-5412. © 2024 IEEE. DOI: 10.1109/ICC51166.2024.10622601. Personal use is permitted, but republication/redistribution requires IEEE permission.}
}
\maketitle
\begin{abstract}
This paper addresses how to handle slice resources for 6G networks at different time scales in an architecture based on an open radio access network (ORAN). The proposed solution includes artificial intelligence (AI) at the edge of the network and applies two control-level loops to obtain optimal performance compared to other techniques. The ORAN facilitates programmable network architectures to support such multi-time scale management using AI approaches. The proposed algorithms analyze the maximum utilization of resources from slice performance to take decisions at the inter-slice level. Inter-slice intelligent agents work at a non-real-time level to reconfigure resources within various slices. Further than meeting the slice requirements, the intra-slice objective must also include the minimization of maximum resource utilization. This enables smart utilization of the resources within each slice without affecting slice performance. Here, each xApp that is an intra-slice agent aims at meeting the optimal quality of service (QoS) of the users, but at the same time, some inter-slice objectives should be included to coordinate intra- and inter-slice agents. This is done without penalizing the main intra-slice objective. All intelligent agents use deep reinforcement learning (DRL) algorithms to meet their objectives. We have presented results for enhanced mobile broadband (eMBB), ultra-reliable low latency (URLLC), and massive machine type communication (mMTC) slice categories.

\end{abstract}

\begin{IEEEkeywords}
DRL, ORAN, Slicing, RRM, eMBB, URLLC, mMTC, QoS, and KPI
\end{IEEEkeywords}

\section{Introduction}
Future wireless networks will connect "everything" (humans, machines, clouds, and servers) and demand uninterrupted high connectivity, data speed, quality, ultra-low latency, and reliability. The beyond 5G networks require dynamic as well as optimal management of resources to fulfill the aforementioned demands. To achieve key performance indicators (KPIs) for applications such as augmented reality (AR), virtual reality (VR), online gaming, vehicular communications, etc., we need to adapt to the new paradigm of machine learning (ML) and AI for wireless communication. In addition to this, traffic demands are increasing exponentially due to technological advancements in recent years, which have to be fulfilled with limited resources. This can be tackled by using ML/AI for resource management between various network components, such as the radio access network (RAN), core network, and to orchestrate it from a broader perspective. The research work presented discusses radio resource management (RRM) for the RAN-edge domain with intelligent orchestration.

The requirements in terms of QoS and service level agreements (SLAs) are getting more stringent day by day. Slicing allows the instantiation of a customized virtual network for each slice on top of the physical network, thus allowing it to meet the requirements of each slice, regardless of how different those requirements are. With enormous datasets, powerful computing, and zero-touch optimization, advanced ML models can offer pervasive, dependable, and almost instant wireless communication for machines and people \cite{survey}. In order to accomplish seamless operation of future wireless networks, network intelligent elements such as on-device distributed learning, deep learning, AI-assisted RAN slicing, edge AI, etc. can be added to various slicing subnets. Several future envisioned applications are categorized into use cases, namely eMBB, URLLC, and mMTC, as defined by 3GPP. For the purpose of achieving the desired KPIs, future research must take into account the aforementioned factors to address a number of issues in these use cases. Hence, we define these 3GPP-based use cases as slice categories to make resource allocation decisions. 

The stringent requirements should be fulfilled by ensuring optimal usage and management of limited resources within a network. The random complexity of the wireless communication environment poses challenges for managing radio resources in the RAN-edge domain. To anticipate the optimal resource allocation for beyond 5G and 6G networks, we need to include AI-based policies. In this direction, several reinforcement learning (RL) techniques have been studied, and DRL techniques such as deep Q networks (DQN) are boosting their appeal to resolve several challenges in wireless communication, including RRM, due to their model-free approach \cite{survey}. The network architecture defined by ORAN standardization facilitates the required integration of such intelligence at the various levels of network components \cite{oran1}. It facilitates AI as a service (AIaaS) using intelligent network components at the edge. The proposed algorithm focuses on optimizing resource allocation, reconfiguration, and utilization. We take advantage of the intelligent aspects of the ORAN network components, such as the near- and non-real-time RAN intelligent controller (near- and non-RT RIC), to manage the resources in an optimal way \cite{oran2}. Proposed algorithms seat at different network components, as xAPPs\footnote[1]{xAPP is the application available at near-RT RIC seating at the edge server as per the ORAN Alliance} and rAPP\footnote[2]{rAPP is the application available at non-RT RIC within slice management and orchestration (SMO) as per the ORAN Alliance}. This paper focuses on managing radio resources at two different levels and time scales. The challenges of efficiently utilizing limited radio resources are tackled in this paper with the following contributions:
\begin{enumerate}
    \item Minimize the maximum utilization of resources at the intra-slice level while ensuring user performance.
    \item Intra-slice reward design helps inter-slice level resource reconfiguration decisions by including the resource utilization $U^{max}_s$ metric and its deviations in the optimization target. It inherently helps inter-slice agents take better decisions at higher centralized level of the architecture.
    \item The state and action spaces of the proposed inter-slice DQN agent are independent of the number of users in the underlying network; hence, the learned optimal decisions or policies remain valid irrespective of the change in scenario.
\end{enumerate}

\section{Related Work}
RAN resource allocation for different slices can be tuned at intra- and inter-slice levels. It corresponds to different timescales in ORAN architecture, referring to near- and non-RT RIC components for smaller and longer time scales, respectively. The state-of-the-art work on intra-slice resource allocation is carried out in \cite{Melike1, UA, UA2, mypaper}. In their study, \cite{Melike1} provide a team learning algorithm designed to improve network performance in an ORAN environment. The proposed method aims to boost collaboration among xAPPs, hence optimizing network efficiency. The reward function being presented is explicitly specified as the aggregate throughput of every individual base station. The main objective is to maximize the aggregate transmission rate across all base stations. The study in \cite{UA} examines the optimization problem pertaining to the combined user association and bandwidth allocation in heterogeneous networks (HetNets). The authors provide an algorithm that incorporates several simultaneous deep neural networks (DNNs) to generate user association solutions. The optimization target for both \cite{UA, UA2} is based on maximization of data rate. It is further analyzed and reduced to define reward functions. A network-slicing 5G architecture and framework while improving QoS is discussed in \cite{mypaper1} for integrating vertical knowledge into existing 5G solutions.

\cite{Melike2, twc} papers talk about inter-slice resource allocation, where \cite{Melike2} proposes a correlated Q-learning-based resource allocation (COQRA) scheme, and the reward function design is based on higher throughput and lower queuing delay for the eMBB and URLLC slices, respectively. The study in \cite{twc} addresses the challenge of restricted radio resources in scenarios where QoS intents need to be prioritized and fulfilled. The reward values are based on metrics like served throughput, average buffer latency, and packet loss rate. The objective of the soft actor-critic (SAC) RL agent is to maximize the reward function values, thereby maximizing the fulfillment of slice intents. Further, \cite{xavier2} discusses the integration of an AI-aided vRAN resource orchestrator into the ORAN architecture, providing interesting details on near-RT and non-RT control loops as well as a variety of deployment scenarios. It highlights the use of AI/ML models for automated resource control and traffic classification, where policies are communicated through interfaces such as A1, O1, and E2. It talks about radio resource management, reward function design, rAPP, xAPP services, etc. The paper emphasizes the importance of AI and ML in next-generation RANs and the flexibility of the O-RAN architecture in supporting different deployment models and functional splits.

The inter-slice decision for resource reconfiguration suffers from a lack of understanding of how the resources are used inside each slice. The presented work addresses this issue. The intelligent agents can learn optimal decisions by rewarding or punishing the network based on its underlying performance as well as resource utilization. This paper proposes a formulation to utilize the DQN technique at the inter- and intra-slice levels. The DQN agents define time-out thresholds for the users inside each slice to schedule the traffic and minimize its maximum resource utilization. Whereas, DQN agents at the inter-slice level take advantage of defined parameters and slice performance to reconfigure resources. 
\section{Network Architecture}
\begin{figure*} [htbp]
        \centering
        \includegraphics[width=16cm, height=6.5cm]{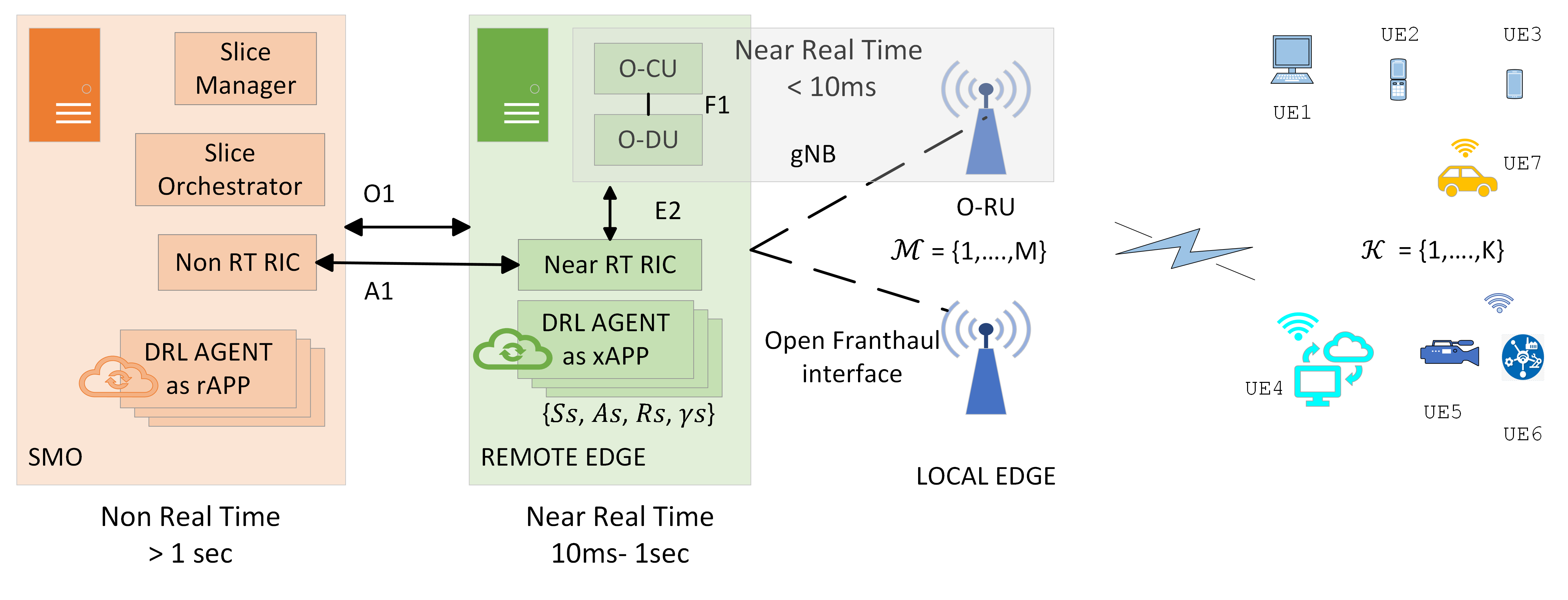}
	\caption{ORAN based network architecture and system model}
	\label{fig:na}
\end{figure*}                                                      
As shown in Figure \ref{fig:na}, we consider ORAN-based architecture. The xAPPs at the remote edge shares information and parameters with the rAPPs seating at the SMO via the A1 interface \cite{oran3}. The SMO and other remote edge network components, including the ORAN centralized unit (OCU) and ORAN distributed unit (ODU), exchange information via O1. ORAN radio units (ORUs) are connected to ODU via an open front-haul link, which can be CPRI or eCPRI. The OCU and ODU can exchange information via F1-CP and F1-UP, whereas OCU exchanges parameters with near-RT RIC via the E2 interface as defined in ORAN specifications.

\section{System Model \& Problem Formulation}
We consider the ORAN network with a general set of users $\mathcal{K} = \{ 1, \cdots, K \}$ and a set of ORUs denoted by $\mathcal{M} = \{ 1, \cdots, M \}$ to serve these users, as shown in Figure \ref{fig:na}. The users request services under one of the slice categories. Each slice user is denoted as $\mathcal{K}_s$, where $s \in \{E,U, M\}$ stands for eMBB, URLLC, and mMTC, respectively. Also, $\mathcal{K} = \mathcal{K}_{E} \cup \mathcal{K}_{U} \cup \mathcal{K}_{M}$ and $\mathcal{K}_{E} \cap \mathcal{K}_{U} \cap \mathcal{K}_{M} = \emptyset $. Here, each ORU can serve all three slices. Let $\mathcal{N} = \{ 1, \cdots, N \}$ be the set of RBs available at each ORU. We take into account frequency division duplexing (FDD) transmission across the downlink (DL). Let the binary variable $a_{k,m,n}$ indicate the association between ORU-users-RB in DL and can be expressed as matrix $A$ where $k \in \mathcal{K}, m \in \mathcal{M} \ \text{and} \ n \in \mathcal{N}$. Every $k$-th user will be associated with one of the ORUs for transmission with $N_{k,m}$ RBs at a given time in the DL. These RBs are grouped as per 5G new radio (NR) into resource block groups (RBGs). We consider numerology $\mu = 0$ as per 5G NR for all users in the network. Each ORU has a fixed bandwidth assigned for each slice category, referred to as $W_{m,s}$. Hence, each $m$-th ORU has a specific number of RBs available under each slice category. We assume additive white gaussian noise (AWGN) with 0 mean and $\sigma^2$ variance. All users in the network have the same speed $V$ and all users follow a random mobility model. We assume a frequency-selective flat fading channel with coefficient $h_{k,m}$ from ORU $m$ to user $k$ for DL. The SINR of the $k$-th user with the $m$-th ORU for DL is expressed as below, given that $P_m$ is the total transmit power of the $m$-th ORU.
\begin{equation}
    \gamma_{k,m,n}=\frac{P_{m,n}\left|h_{k,m,n}\right|^2}{I_{k,m,n}+\sigma^2} 
\end{equation}
Where the interference from the $m^-$ ORUs (other than the $m$-th ORU) for the $k$-th user can be formulated as,
\begin{equation}
    I_{k,m,n}=\sum_{m^-\in{\mathcal{M}\setminus m}}{P_{m^-,n}\left|h_{k,m^-,n}\right|^2}   
\end{equation}
Each user has QoS requirements expressed in terms of the minimum data rate $R_k^{\text{min}}$ and the maximum allowed delay $d_k^{\text{max}}$ where $k \in \mathcal{K}$. We assume that all users from the same slice have the same QoS requirements. We consider the eMBB, URLLC, and mMTC users follow a periodic deterministic data traffic model with fixed packet arrival intervals defined per slice category.

\subsection{Intra-slice Resource Allocation}
A fixed number of RBGs are available at each slice, as informed by SMO. These RBGs are denoted by $Z_s$, where $s \in \{E,U,M\}$ and contain $N$ number of RBs. The proposed intra-slice intelligent agent helps to take a decision for allocating a number of RBs to end users under the slice $s$ within available resources. 
The intra-slice level scheduler works based on a timeout threshold $\tau_{th}^s$ to schedule packets for all users under a slice. The ORU assigns RBs to the user whose $l$-th packet has the lowest difference between the current delay $d_{k,l}$ and $\tau_{th}^s$, where $l \in L_k$ is the number of packets to be transmitted at user $k$. Once the resources are allocated by the scheduler, we calculate the maximum utilization of the resources, their deviation, and the penalties experienced by packets due to the expiration of the QoS maximum delay limit. The maximum utilization is defined below and calculated for the total number of resources utilized by all users under a specific slice.
\begin{equation}
    U_{s,t} = \sum_{k \in \mathcal{K}_s} \sum_{m \in \mathcal{M}} N_{k,m,t} \cdot S \cdot l \cdot O_{k,m,t} \ ,\  U^{max}_{s,t} =  \max_{t \in TTI} U_{s,t}
\end{equation}
Where $S$ is the number of symbols, $l$ is the number of spatial multiplexing layers, and $O_{k,m,t}$ is the modulation order. The deviations of $U^{max}_{s,t}$ are observed and forwarded to inter-slice resource allocation for better reconfiguration decisions. The deviation is calculated as below:
\begin{equation}
\begin{split}
    \delta_s = \frac{U^{max}_{s,t} - \mu_s}{\mu_s}  \ \ \text{where,} \ \
    \mu_s = \frac{1}{|T|} \sum_{t = 0}^{T} U^{max}_{s,t}
\end{split}
\end{equation}
The expiration penalty is defined as the number of packets per user that are transmitted exceeding the QoS threshold, such as,
\begin{minipage}{0.23\textwidth}
\begin{equation}
\begin{split}
    \epsilon_{k,l} &= 1 \ \ \text{if} \ d_{k,l} \geq d_k^{\text{max}} \\
    &= 0 \ \ \text{otherwise} 
    \end{split}
\end{equation}
\end{minipage}
\hfill
\begin{minipage}{0.23\textwidth}
\begin{equation}
    \epsilon_{s,t} = \sum_{k \in \mathcal{K}_s} \sum_{l \in L_k} \epsilon_{k,l}
\end{equation}
\end{minipage}

\subsection{Performance and Objective}
The performance of a user or a slice is evaluated with metric throughput and delay. The average throughput per slice $R^s_{avg}$ is calculated as the sum of the throughput for all users divided by the number of users in that slice. We normalize the individual user throughput by dividing it by the minimum QoS requirement. The average throughput over the $t$ transmission time intervals (TTIs) for user $k \in \mathcal{K}$ can be expressed as,
\begin{equation}
\begin{split}
    R_{k,m,t} &= N_{k,m} \times W_{RB} \ \log_2(1+\gamma_{k,m,n}) \\
    {\bar{R}}_{k} &= \frac{\sum_{t\ \in T T I} R_{k,m,t}}{|TTI|}
\end{split}
\end{equation}
Where $|\cdot|$ is the set cardinality operator, $W_{RB}$ is the bandwidth of a single RB, and $N_{k,m}$ is the number of resource blocks allocated to user $k$ at time $t$. Therefore, the average normalized throughput of the individual user is given as below:
\begin{equation}
    {\bar{R}}_{k,norm}= \frac{{\bar{R}}_k}{R_k^{min}}
\end{equation}
Therefore, average slice throughput performance is calculated as,
\begin{equation}
     R_{avg}^s =\frac{\sum_{k \in \mathcal{K}_s} {\bar{R}}_{k,norm}}{|\mathcal{K}_s|} =\frac{1}{|\mathcal{K}_s|}\sum_{k \in \mathcal{K}_s}\frac{{\bar{R}}_k} {R_k^{min}} 
\end{equation}
The total delay experienced per user is a combination of delays experienced by each packet in queue, transmission, and processing, as given below, where sub-index $l$ is the packet index for the user. The total delay per user is calculated as the average delay experienced by all packets.
\begin{equation}
    {d}_{k,l} =\ d_{k,l}^{tx}+d_{k,l}^{que}+d_{k,l}^{processing} \ \text{and} \
    {\bar{d}}_k =\ \frac{\sum_{l\ \in L} d_{k,l}}{\left|L\right|} 
\end{equation}
The normalized delay is defined as,
\begin{equation}
    {\bar{d}}_{k,norm}=\ \frac{{\bar{d}}_k}{d_k^{max}}
\end{equation}
Hence, average user delay performance per slice is calculated as,
\begin{equation}
     d_{avg}^s =\frac{\sum_{k \in \mathcal{K}_s} {\bar{d}}_{k,norm}}{|\mathcal{K}_s|} =\frac{1}{|\mathcal{K}_s|}\sum_{k \in \mathcal{K}_s}\frac{{\bar{d}}_k} {d_k^{max}} 
\end{equation}
The performance indicator for user $k\in \mathcal{K}_s$ is defined as the number of users achieving average normalized throughput and average normalized delay greater than or equal to one, given the total number of users in DL. It is expressed as,
\begin{equation}
    \begin{split}
        {P}^{QoS}_k\ &=1\ if\ \left\{{\bar{R}}_{k,norm}\geq1 \land {\bar{d}}_{k,norm}\geq1\right\} \\
        {\bar{P}}_s &=\frac{|{P}^{QoS}_k|}{|\mathcal{K}_s|}
    \end{split}
\end{equation}
The objective of different slices in this proposal is to minimize the maximum utilization of resources within each slice while achieving predefined QoS performance. It is expressed as below:
\begin{equation}
\begin{split}
    &\min U^{max}_{s,t} \\
    \text{subjected to:} \ \ &{P}^{QoS}_k =1 , \ k \in \mathcal{K}_s
\end{split} 
\end{equation}

\section{Proposed multi timescale intelligent agents using DRL}
All intelligent agents use the DRL technique, namely DQN, with similar objective functions at the intra-slice level to minimize the maximum utilization of resources. Here, intra-slice intelligent agents schedule the available slice resource and allocate the number of RBs to end users in such a way that the performance of the slice is maintained according to QoS requirements while trying to minimize the utilization of resources in a single TTI. Whereas, inter-slice agents reconfigure the fixed RBG pool within different slices based on slice performance KPIs and the resource utilization information referred to by underlying intra-slice agents. The reconfiguration decisions are then forwarded via A1 to intra-slice intelligent agents. In this study, distinct intelligent agents for the slice types, eMBB, URLLC, and mMTC, are proposed. 

Based on the state of the network environment at any given time, the DQN determines the threshold level $\tau_{th}^s$ for each slice and evaluates the optimum course of action. The DQN trains and learns these values for various network environments. The formulated solutions in the earlier section take into account the $\tau_{th}^s$ and evaluate a decision for RB allocation to serve the corresponding end user. The proposed Markov decision process (MDP) for intelligent agents not only ensures better performance for the KPIs compared to baseline but also minimizes the utilization of resources within slices. 
\subsection{MDP for Intra-slice Intelligent Agents}
The Markov Decision Process (MDP) for the intelligent agents based on the above formulation is defined with a tuple of $\left\{S_s,A_s, R_s\ ,\Gamma_s\right\}$ corresponding to state, action, reward, and discount factor, where subscript $s$ indicates slice type for intelligent agents $s\in\{E,U, M\}$. The state space $S_s$ of the environment includes the current log normalized channel matrix $H_{M\times K}$ and the number of bits to be transmitted in the buffer per user ${\bar{B}}_k$. The state space definition remains the same for the different slices, as it represents the current state of the environment based on which the actions will be chosen by the intelligent agent. Different intelligent agents learn different timeout thresholds that are specific and valid to the varying traffic flow for the services and users under each slice.
\begin{equation}
    S_s =\left\{H_{M\times K},{\bar{B}}_k\right\}  \ \text{for} \ k, m\in \mathcal{K}_s, \mathcal{M}_s
\end{equation}
The action space is discretized timeout threshold values. The set of $\text{D}^{s}$ discrete values ranging from $\tau_{\text{min}}$ to \(\tau_{\text{max}}\) can be expressed as:
\begin{equation}
    A_s = \tau^{s}_{th} \in \{\tau^{s}_{\text{min}}, \tau^{s}_{\text{min}} + \tau^{s}_{step}, \ldots, \tau^{s}_{\text{max}} - \tau^{s}_{step}, \tau^{s}_{\text{max}}\}
\end{equation}
The action space is bounded, hence DQN can converge with limited number of actions.
The reward function is designed as shown below. It is a combination of three crucial parameters: maximum utilization of resources in the slice, slice throughput performance, and expiration penalty. All these three parameters are weighted with $\alpha, \beta, \text{and} \ \gamma$, respectively. These weighted values are set differently for each slice according to SLAs. The reward function can be expressed as follows:
\begin{equation}
        {R}^i_s\left(S_s,A_s\right) = \alpha_s \cdot U^{max}_{s,i}+ \beta_s \cdot R_{avg}^{s,i} - \gamma_s \cdot \epsilon_{s,i}
\end{equation}
Where i indicates the episode or run of the DQN algorithm and $s \in\{E,U, M\}$.
\subsection{MDP for Inter-slice Intelligent Agents}
A tuple of $\left\{S,A,R\ ,\Gamma\right\}$ corresponding to state, action, reward, and discount factor represents MDP for inter-slice intelligent agent using DQN. The state space is defined as shown below:
\begin{equation}
    S =\left\{R_{avg}^{s}, U^{max}_{s}, \delta_s, \epsilon_{s}\right\} \ \text{for}, s\in \{ E, U, M\}
\end{equation}
The action space for this intelligent agent is defined as the combination of all available RBGs distributed among slices. The number of actions varies based on the total number of RBGs available at the SMO $Z$ and the number of slices $s$. For the proposed study, we analyze three slices, and the combinations for RBGs are indicated with $Z^{comb}$. Therefore, the action space has $D$ such combinations as indicated below:
\begin{equation}
    A =  \{Z^{comb}_1, \ldots , Z^{comb}_D\}
\end{equation}
Where $Z^{comb}_D = \{Z_E, Z_U, Z_M\}$ shows the corresponding number of $Z_s$ RBGs allocated to $s$ slice. Further, the reward is defined as a combination of slice performance metrics, as given below:
\begin{equation}
        {R}^i\left(S,A\right) = \sum_{s} \left( R_{avg}^{s, i} -  d_{avg}^{s, i} \right) %
\end{equation}
\subsection{DQN}
We use the DQN method from DRL algorithms. The Q value, target Q value, and loss function are calculated as given below \cite{book2}.
\begin{equation}
    L_s=\left[{TD}_s-\left(Q^i_s\left(s_s,a_s\right)\right)^2\right]
\end{equation}
\begin{equation}
     Q^i_s\left(s_s,a_s\right) =\ Q^i_s\left(s_s,a_s\right)+\Omega_s\times L_s
\end{equation}
Here, $s \in \{E,U, M\}$ and the values for learning rate $\Omega_s$, discount factor $\Gamma_s$ as well as other hyper-parameters are indicated in the simulation setup. The temporal difference, or target Q value, is calculated as follows:
\begin{equation}
    {TD}_s={r_s}^{i+1}+\Gamma_s\ \max_{\substack{a_s\in A_s}} {Q^{i+1}_s\left(s_s,a_s\right)}
\end{equation}
\subsection{Discussion on Complexity Analysis}
The proposed solution is scalable. Given the increased number of users as well as ORUs in the scenario, the state and action space of the inter-slice agent remain unaffected. The same agent could be utilized.  Whereas, for an intra-slice agent, the 
state space will have a significant increase. This issue could be tackled by increasing the number of slices and instantiating an additional agent. Now, an increase in the number of slices would increase the state space of the inter-slice agent very slowly without a higher impact on complexity.
\section{Simulation and verification}
For simulation setup with 5G NR $\mu=0$, we set sub-carrier spacing to 15 kHz and 12 subcarriers. Hence the bandwidth of a single RB, $W_{RB}=180kHz$ and one RBG contains 6 RBs. We consider total $Z=14$ RBGs available to distribute within three slices. The simulation and verification set up for transmission power, height and placement of ORUs, noise figure, variance, and pathloss exponent are as per tables A1-12 to A1-23 of the ITU-R recommendations and are in compliance with ORAN standards \cite{itu1,itu2,oran2}. We consider a total of 3 ORUs to be available, with 3 slices, and a total of 9 users, with 3 users per slice. Other simulation parameters are mentioned in Table \ref{tab:Param}. The QoS requirement thresholds are set by considering average expected KPIs under each slice category \cite{3gpp}. 
The simulation time is set to 50 seconds. The intra-slice DQN agents run every 10 TTIs, whereas the inter-slice agents run every 200 TTIs to obtain results. Furthermore, the inter-slice agent can be set to run every 1 second if it is trained prior for a minimum of 500 seconds.
\begin{table}[htbp]
\centering
\caption{Simulation Parameters}
\label{tab:Param}
\begin{tabularx}{\linewidth}{X X X}
\toprule
Type & Parameter & Value  \\
\midrule
QoS Requirements &  $\bm{R}_k^{\text{min}}$ (Mbps)& E=16, U=3.8, M=0.5 \\
&  $\bm{d}_k^{\text{max}}$ (ms)& E=10, U=2, M=20  \\
\midrule
Packet Generation & Arrival Interval (ms) & E=0.5, U=1, M=0.5 \\
& Packet Size (Bytes) & E=1024,U=480,M=32 \\
\midrule
DRL agent & N/W Arch & 64 $\times$ 64 $\times$ 256 \\ 
Intra-slice& $\Omega$ & $1 \times 10^{-3}$ \\ 
& Batch size & 256 \\ 
\midrule
Inter-slice & N/W Arch & 256 $\times$ 256 \\ 
& $\Omega$ & $1 \times 10^{-4}$ \\ 
& Batch size & 64 \\ 
\midrule
& TN Update & 100 \\ 
\bottomrule
\end{tabularx}
\end{table}
\vspace{-0.7mm}
\section{Results}
\begin{figure}[htbp]
    \centering
    \begin{subfigure}{0.46\textwidth}
        \centering
        \includegraphics[width=\textwidth]{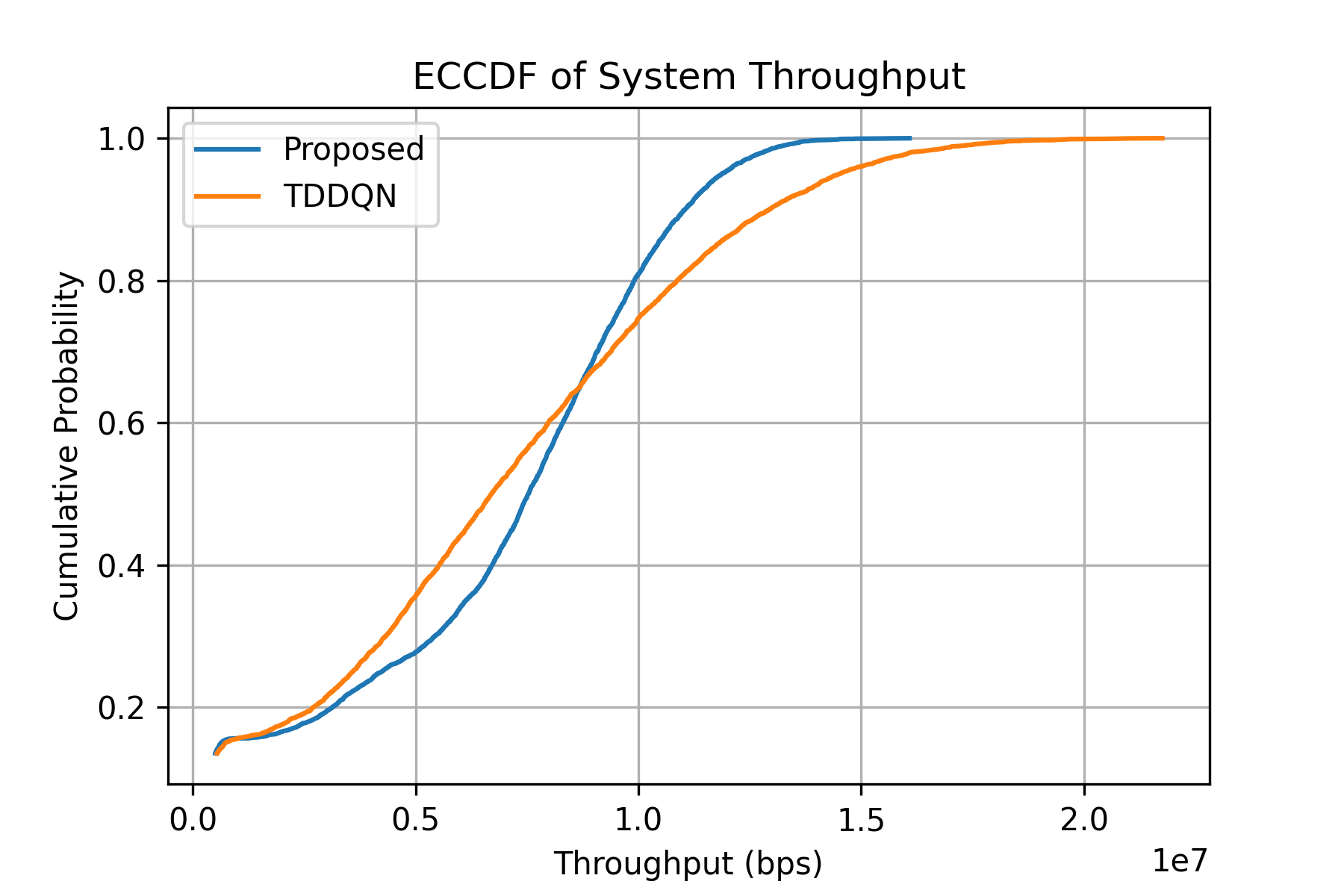}
        \caption{System Throughput ($R^s$) Performance Comparison} %
        \label{subfig:subfigure1}
    \end{subfigure}
    \hfill
    \begin{subfigure}{0.46\textwidth}
        \centering
        \includegraphics[width=\textwidth]{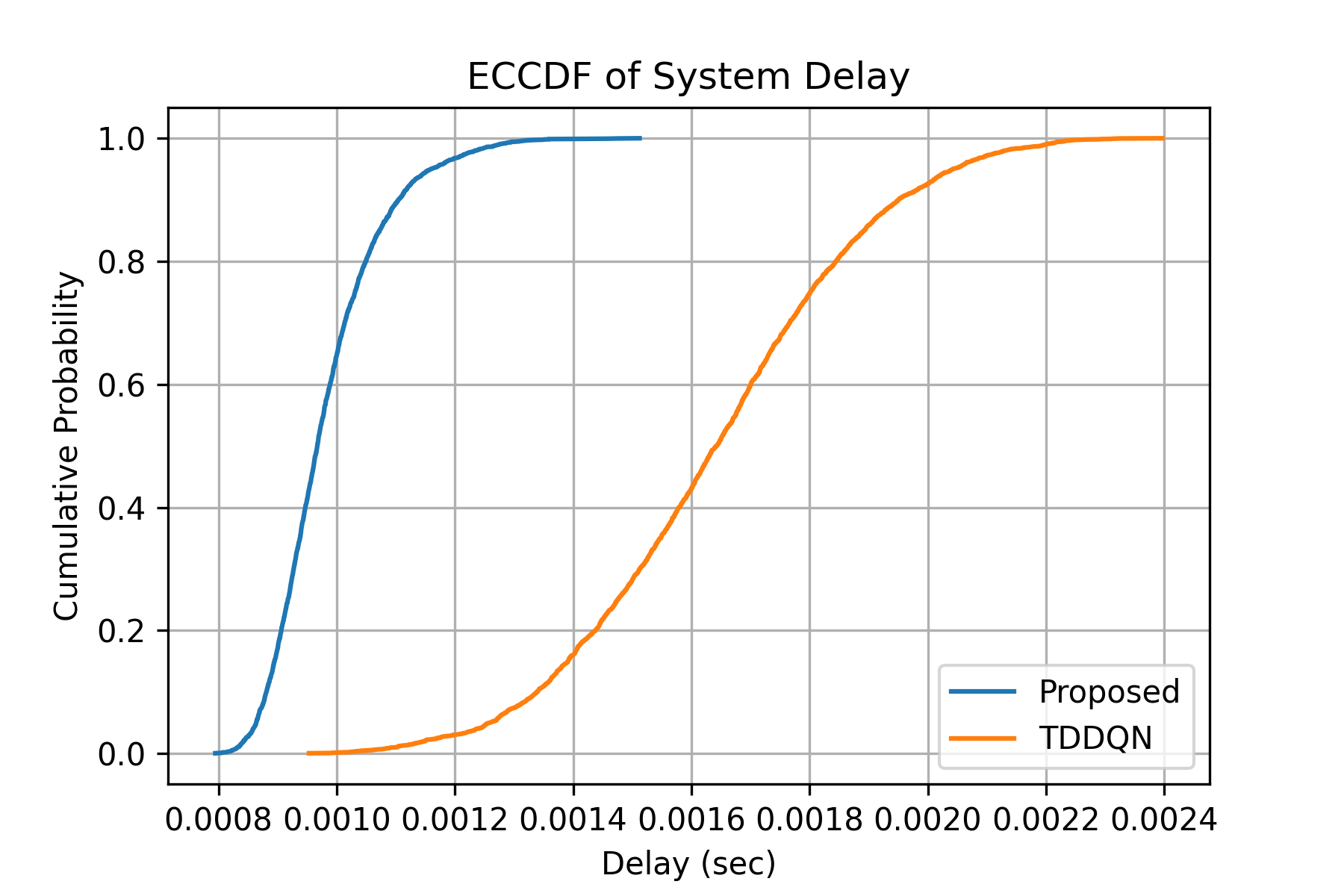}
        \caption{System Delay ($d^s$) Performance Comparison}
        \label{subfig:subfigure2}
    \end{subfigure}
    \caption{System KPIs}
    \label{fig:R1}
\end{figure}

\begin{figure}[htbp]
    \centering
    \includegraphics[width=0.46\textwidth]{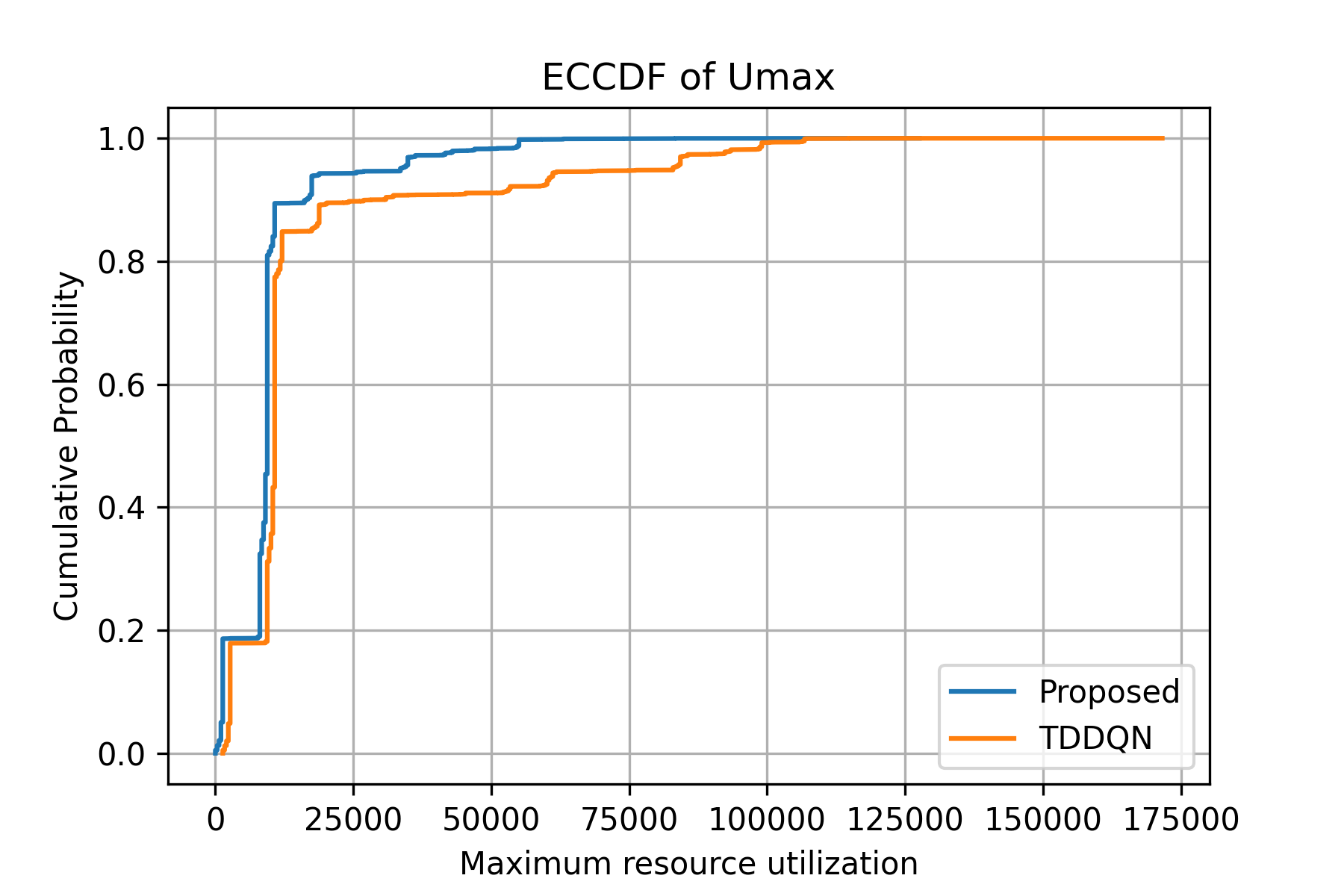}
    \caption{Resource Utilization ($U^{max}$) comparison}
    \label{fig:R2}
\end{figure}

The results are obtained with the simulation setup discussed in the above section. The weights used in reward functions for each slice as defined in equation (17) and other parameters to define action space for intra-slice agents are given in Table \ref{tab:Param_result}. The length of the action space for all intra-slice agents depends on $D^s$ and is set to 20. Whereas, for an inter-slice agent, the length of action space is 120 based on a possible combination of the available 14 RBGs assigned between 3 slices.
Intra-slice intelligent agents achieve convergence within 180–250 DQN runs. Furthermore, the offline 100-second-trained inter-slice intelligent agent achieves convergence for the reward function within the first 50 runs. 
\begin{table}[htbp]
\centering
\caption{Intra-slice agent parameters}
\label{tab:Param_result}
\begin{tabular}{l l l l}
\toprule
Parameter & eMBB & URLLC & mMTC  \\
\midrule
$\tau_{\text{min}}$ (ms) & 2 & 0.5 & 5 \\
$\tau_{\text{max}}$ (ms) & 10 & 2 & 15\\
$\alpha$ & 0.3 & 0.3 & 0.3 \\
$\beta$ & 0.4 & 0.3 & 0.35 \\
$\gamma$ & 0.3 &  0.4 & 0.35\\
\bottomrule
\end{tabular}
\end{table}
Subfigure~\ref{subfig:subfigure1} and \ref{subfig:subfigure2} show the system KPI performance comparison for throughput and delay. The Empirical Cumulative Distribution Function (ECCDF) is plotted to compare performance between the proposed scheme and throughput-delay-based DQN, further referred to as TDDQN. The TDDQN uses a similar simulation setup, except it does not include maximum utilization of resources in its reward function; instead, it is defined as other state-of-the-art techniques to maximize the throughput same as equation (6) from work conducted in \cite{Melike2}. 
The result shows that the minimum system delay is reduced by 5–12\% in each slice as the proposed intra-slice agents define a separate timeout threshold. The proposed algorithms result in reduction of the maximum utilization of resources by 34\%, as shown in Figure \ref{fig:R2} compared to TDDQN while maintaining the overall system performance. This helps with better resource reconfiguration at the inter-slice level.
\section{Conclusion and Future Scope}
The proposed scheme is crucial for beyond 5G networks due to increasing traffic demands to be served within limited resources. Including knowledge of resource utilization can help the network manage the limited resource pool optimally without degrading performance, as explained in the result section. The proposed scheme can be extended to define different timeout thresholds based on user priorities within each slice category. This will improve intra-slice resource scheduling, leading to better KPIs with variation in SLAs. Intra- and inter-slice agent's initial hidden layers learn similar knowledge that can be transferred to achieve quicker convergence. 

\section*{Acknowledgment}
The conducted work is part of MSCA ITN SEMANTIC(861165), H2020-5GPP MARSAL(101017171), SNS-JU ADROIT-6G(101095363), ELIDEK ENABLE-6G(016294), RF-VOLUTION(PID2021-122247OB-I00), \& Generalitat de Catalunya Grant 2021 SGR 174.


\end{document}